\UseRawInputEncoding
\documentclass[preprints,article,accept,moreauthors,pdftex]{Definitions/mdpi} 

\firstpage{1} 
\pubvolume{xx}
\issuenum{1}
\articlenumber{5}
\pubyear{2020}
\copyrightyear{2020}
\history{}
\Title{Interferometry in an Atomic Fountain with Ytterbium Bose-Einstein Condensates }

\Author{Daniel Gochnauer $^{1}$, Tahiyat Rahman $^{1}$, Anna Wirth-Singh $^{1}$, and Subhadeep Gupta $^{1}$*}

\AuthorNames{Daniel Gochnauer, Tahiyat Rahman, Anna Wirth-Singh, and Subhadeep Gupta}

\address{%
$^{1}$ \quad Department of Physics, University of Washington, Seattle, Washington 98195, USA\
}

\corres{Correspondence: deepg@uw.edu}
\abstract{We present enabling experimental tools and atom interferometer implementations in a vertical ``fountain'' geometry with ytterbium Bose-Einstein condensates. To meet the unique challenge of the heavy, non-magnetic atom, we apply a shaped optical potential to balance against gravity following evaporative cooling and demonstrate a double Mach-Zehnder interferometer suitable for applications such as gravity gradient measurements. Furthermore, we also investigate the use of a pulsed optical potential to act as a matter wave lens in the vertical direction during expansion of the Bose-Einstein condensate. This method is shown to be even more effective and results in a reduction of velocity spread (or equivalently an increase in source brightness) of more than a factor of five, which we demonstrate using a two-pulse momentum-space Ramsey interferometer. The vertical geometry implementation of our diffraction beams ensures that the atomic center of mass maintains overlap with the pulsed atom optical elements, thus allowing extension of atom interferometer times beyond what is possible in a horizontal geometry. Our results thus provide useful tools for enhancing the precision of atom interferometry with ultracold ytterbium atoms.}

  % List three to ten pertinent keywords specific to the article, yet reasonably common within the subject discipline.

\begin{document}
%%%%%%%%%%%%%%%%%%%%%%%%%%%%%%%%%%%%%%%%%%

\section{Introduction}
Pulsed optical lattices are crucial tools for high precision atom interferometry (AI), with applications ranging from tests of fundamental physics to force sensing \cite{bouc11,mazz15,hupo17,park18,asen17}. AI in free space rather than in a trapped geometry has the inherent advantage of not being susceptible to systematic effects from the confining potentials. Terrestrial pulsed-lattice atom interferometers have relied on a vertical geometry of diffraction beams in order to fully realize the inherent power of the method, as the loss of spatial overlap with the pulsed lattice from atoms falling due to gravity is suppressed in this configuration. 

While laser cooled atoms have found pronounced success as sources for precision AI, a Bose-Einstein condensate (BEC) source offers improvement with an inherently narrow velocity distribution, which increases the coherence length and allows for longer interferometer times due to the slow spread of the atomic spatial distribution during free expansion. Spin-singlet ground state atoms, such as Sr and Yb, are particularly appealing for AI because of their near-insensitivity to external magnetic fields and the availability of several narrow optical transitions \cite{jami14,mazz15,rudo20}. Furthermore, the heavy nucleus of the Yb atom supports the stability of multiple isotopes allowing for systematic cross-checks within the same apparatus as well as the possibility to perform AI with degenerate Bose or Fermi gases. 

Prior work with Yb BEC atom interferometers \cite{jami14,plot18,goch19,mcal20} have all been restricted to geometries with horizontally oriented laser beams as the atom-optic elements. Adapting to a vertical geometry poses significant challenges due to the larger velocity spread in the vertical direction for BECs expanding out of typical atom traps. In this paper, we investigate solutions to these challenges using two separate methods of atom manipulation via light shifts from laser beams shaped in position and in time. Using these methods we successfully demonstrate AI with Yb BECs in a vertically oriented atomic fountain geometry and establish tools for future enhancement of AI precision. 

The rest of this paper is organized as follows. In Section 2, we describe the production of our BEC source and the vertical fountain launch which prepares the atoms for interferometry. In Section 3, we present a gravity compensation beam method for reducing vertical velocity spread. With this method in place, we report on the first vertical Yb interferometer, performed in a double Mach-Zehnder configuration, in Section 4. Finally, in Section 5, we implement delta-kick cooling as an alternative method to the gravity compensation beam, and we demonstrate effectiveness of the technique through coherence length measurements using momentum-space Ramsey interferometry. 

%Citing a journal paper \cite{ref-journal}. And now citing a book reference \cite{ref-book}. Please use the command \citep{ref-journal} for the following MDPI journals, which use author-date citation: Administrative Sciences, Arts, Econometrics, Economies, Genealogy, Humanities, IJFS, JRFM, Languages, Laws, Religions, Risks, Social Sciences.
 
%%%%%%%%%%%%%%%%%%%%%%%%%%%%%%%%%%%%%%%%%%
\section{Yb BEC Fountain}

We first briefly describe Yb BEC production in our apparatus \cite{plot18thesis,plot18} and then present the launch process which initiates the atomic fountain.

%%%%%%%%%%%%%%%%%%%%%%%%%%%%%%%%%%%%%%%%%%
\subsection{BEC Source and Atom Optics}
%\unskip
%\subsubsection{Subsubsection}

%Bulleted lists look like this:
%\begin{itemize}[leftmargin=*,labelsep=5.8mm]
%\item	First bullet
%\item	Second bullet
%\item	Third bullet
%\end{itemize}

%Numbered lists can be added as follows:
%\begin{enumerate}[leftmargin=*,labelsep=4.9mm]
%\item	First item 
%\item	Second item
%\item	Third item
%\end{enumerate}

%The text continues here.
Each of the experiments reported in this work begins with a trapping and cooling sequence for the production of a ytterbium (${}^{174}{\rm Yb}$) Bose-Einstein condensate consisting of $10^5$ atoms \cite{plot18thesis}. A Yb atomic beam emerging from an effusive oven is slowed in a first stage through an increasing-field Zeeman slower and in a second stage using a pair of crossed laser beams \cite{plot20}. The slowed atoms are then captured in a magneto-optic trap (MOT). The broad ($\Gamma_b=2\pi\times29$~MHz) dipole transition (${}^1S_0\rightarrow{}^1P_1$) at $\lambda_b=399\,$nm is used to slow the atoms, while the narrow ($\Gamma_g=2\pi\times182$~kHz) intercombination transition (${}^1S_0\rightarrow{}^3P_1$) at $\lambda_g=556~\,$nm is used for the MOT trapping beams (see Fig.\ref{fig:Figure1}(a)).

Following cooling in the MOT, atoms are transferred into a crossed optical dipole trap (ODT) for evaporative cooling towards BEC. The ODT is formed by a pair of 532~nm beams: one oriented horizontally, defining the x-axis, and the other nearly along the vertical y-axis. The evaporative cooling stage concludes with a BEC of $10^5$ atoms in harmonic confinement characterized by trap frequencies $\omega_{x,y,z}=2\pi \times (16,200,80)$~Hz. Subsequently, the BEC is released by suddenly switching off the ODT before the application of atom optics pulses. We note that the tightest confinement direction in the ODT is along the vertical axis, in order to counter the gravitational force on the heavy atom. A consequence of this general characteristic of optically trapped Yb is that the expansion of the BEC after release is mostly along the vertical, as most of the initial chemical potential is converted to kinetic energy in this direction \cite{cast96,jami11}, a quantity we measure to be $k_B \times (42\pm5)$~nK through absorption imaging after long expansion times.  

The optical lattice used for the vertical fountain launch and the interferometer optics has a waist of $1.8\,$mm and is composed of a pair of vertically-oriented counter-propagating laser beams that are aligned to the atoms and controlled by independent acousto-optic modulators (AOMs). The optical frequency of the lattice is detuned from the ${}^1S_0\rightarrow{}^3P_1$ transition by $\Delta$ and the frequency difference between the two lattice laser beams is $\delta$ (see Fig.~\ref{fig:Figure1}(a) and inset to 1(b)). For the work presented here, $\Delta$ is set to $+3500\Gamma_g$ ($2\pi \times 637\,$MHz), except in Section 5 where $\Delta = -3500\Gamma_g$. The + sign is depicted in Fig.~\ref{fig:Figure1}(a). The quantity $\delta$ is always less than $2\pi \times 1\,$MHz and varied with sub-Hz precision using direct digital synthesis radio-frequency sources that drive the lattice AOMs.  

We note that the measured kinetic energy in the vertical direction corresponds to a velocity spread of $\Delta v \simeq 0.5 v_{\rm rec}$ where $v_{\rm rec}=\hbar k_g/m$ is the recoil velocity with $k_g=2\pi/\lambda_g$ and $m$ is the mass of a Yb atom. Since the coherence time relevant for signal-to-noise in atom interferometry typically scales with $1/\Delta v$, it is important to address the reduction of this value, and much of the present work demonstrates successful techniques towards this end for Yb BEC vertical interferometers.  

\subsection{Vertical Fountain Launch}

\begin{figure}[H]
	\centering
	\includegraphics[width=1.0\textwidth]{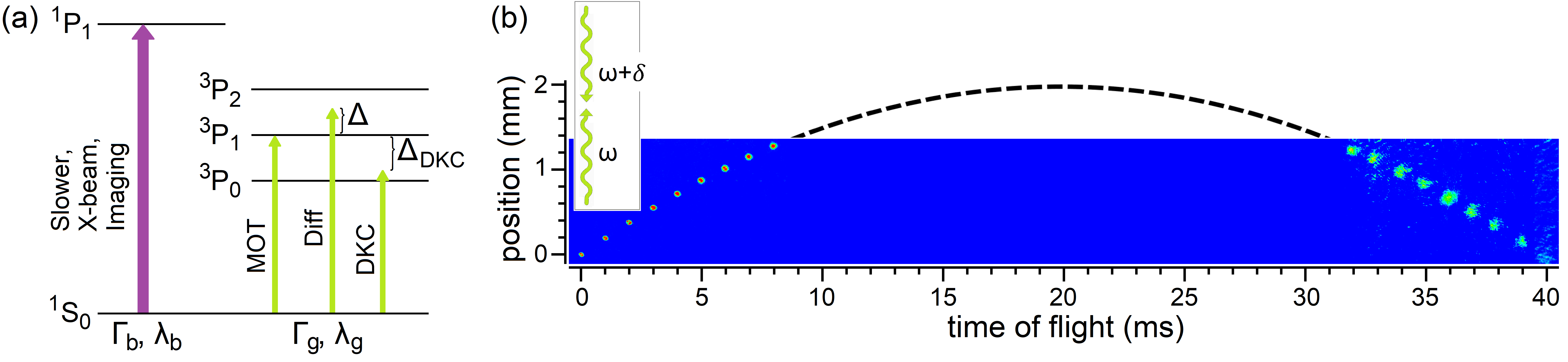}	\caption{(\textbf{a})~Energy level diagram for ytterbium showing the optical transitions used in the experiment.  (\textbf{b})~Demonstration of a vertical fountain launch with 30 ground-band Bloch oscillations. The relative frequency sweep of the lattice beams was $\dot{\delta}=2\pi\times 600$~kHz/ms. Time-of-flight absorption images show the trajectory of the atom cloud at variable times following the launch. The dashed line indicates the part of the trajectory that is out of view of the absorption imaging setup.}
	\label{fig:Figure1}
\end{figure}

The vertical launch is performed by using Bloch oscillations for large momentum transfer. Such processes were carried out in the following sequence: (i)~adiabatically turning on the optical lattice in the frame of the falling atoms, (ii)~chirping the relative frequency difference of the lattice beams, $\delta$, to accelerate the atoms, and (iii)~adiabatically turning off the lattice once the desired atom velocity had been reached. During lattice turn-on and turn-off, an additional chirp of $\dot{\delta}=2gk_g$ was necessary to maintain an inertial frame. Here $g$ is the acceleration due to gravity. The frequency sweep during the acceleration step was uniquely chosen for each set of experiments to optimize momentum transfer efficiency. A representative fountain launch is shown in Fig.~\ref{fig:Figure1}(b).

%All figures and tables should be cited in the main text as Figure 1, Table 1, etc.

%\begin{table}[H]
%\caption{This is a table caption. Tables should be placed in the main text near to the first time they are cited.}
%\centering
%% \tablesize{} %% You can specify the fontsize here, e.g., \tablesize{\footnotesize}. If commented out \small will be used.
%\begin{tabular}{ccc}
%\toprule
%\textbf{Title 1}	& \textbf{Title 2}	& \textbf{Title 3}\\
%\midrule
%entry 1		& data			& data\\
%entry 2		& data			& data\\
%\bottomrule
%\end{tabular}
%\end{table}

%\begin{listing}[H]
%\caption{Title of the listing}
%\rule{\textwidth}{1pt}
%\raggedright Text of the listing. In font size footnotesize, small, or normalsize. Preferred format: left aligned and single spaced. Preferred border format: top border line and bottom border line.
%\rule{\textwidth}{1pt}
%\end{listing}

%% If the documentclass option "submit" is chosen, please insert a blank line before and after any math environment (equation and eqnarray environments). This ensures correct linenumbering. The blank line should be removed when the documentclass option is changed to "accept" because the text following an equation should not be a new paragraph. 

%%%%%%%%%%%%%%%%%%%%%%%%%%%%%%%%%%%%%%%%%%
\section{Gravity Compensation by Shaped Optical Beam}

Atoms trapped in our ODT experience a potential proportional to the shape of the ODT beams as well as gravity. Due to the linear gravitational potential, there is a nonzero minimum allowable depth of the ODT such that it remains a trapping potential. This value increases as the vertical width of the horizontal ODT beam increases, resulting in a conflict between desires for both low trap frequency and low trap depth. This constraint can be lifted, however, by compensating the gravitational potential with an appropriately tuned linear optical potential in the trapping region \cite{shib20}. Importantly, this technique is accessible for all atoms, including non-magnetic atoms such as ytterbium. Using a time-averaging technique discussed below, we implemented an appropriately shaped optical potential $U_S$ to weaken the confinement along the vertical direction and thus the vertical kinetic energy during expansion after release from ODT. Assuming a Gaussian shape of the trapping beam (waist $w_0$), we can write the total potential seen by the atoms as
\begin{eqnarray}
U(y) &=& U_0 e^{-2y^2/w_0^2} - mgy + U_S(y) \label{eq:odt} \\ %%%%%%% U_S instead of U_L ?
	&\simeq& U_0(1-\frac{2y^2}{w_0^2}) + (\alpha -mg)y = U_0 + \frac{1}{2}m\omega_y^2y^2 +(\alpha-mg)y \label{eq:odtHO}
\end{eqnarray}
where $U_0 (<0)$ is the peak ODT Stark shift, $w_0$ is the ODT beam waist, $U_S(y)$ is the additional gravity compensation potential with $\alpha = \partial U_S/\partial y|_{y=0}$, and we have kept lowest order terms around $y=0$ for Eq. \ref{eq:odtHO}. $U_S$ is tuned such that $\alpha \simeq mg$ at the location of the atoms.

\begin{figure}[H]
	\centering
	\includegraphics[width=0.70\textwidth]{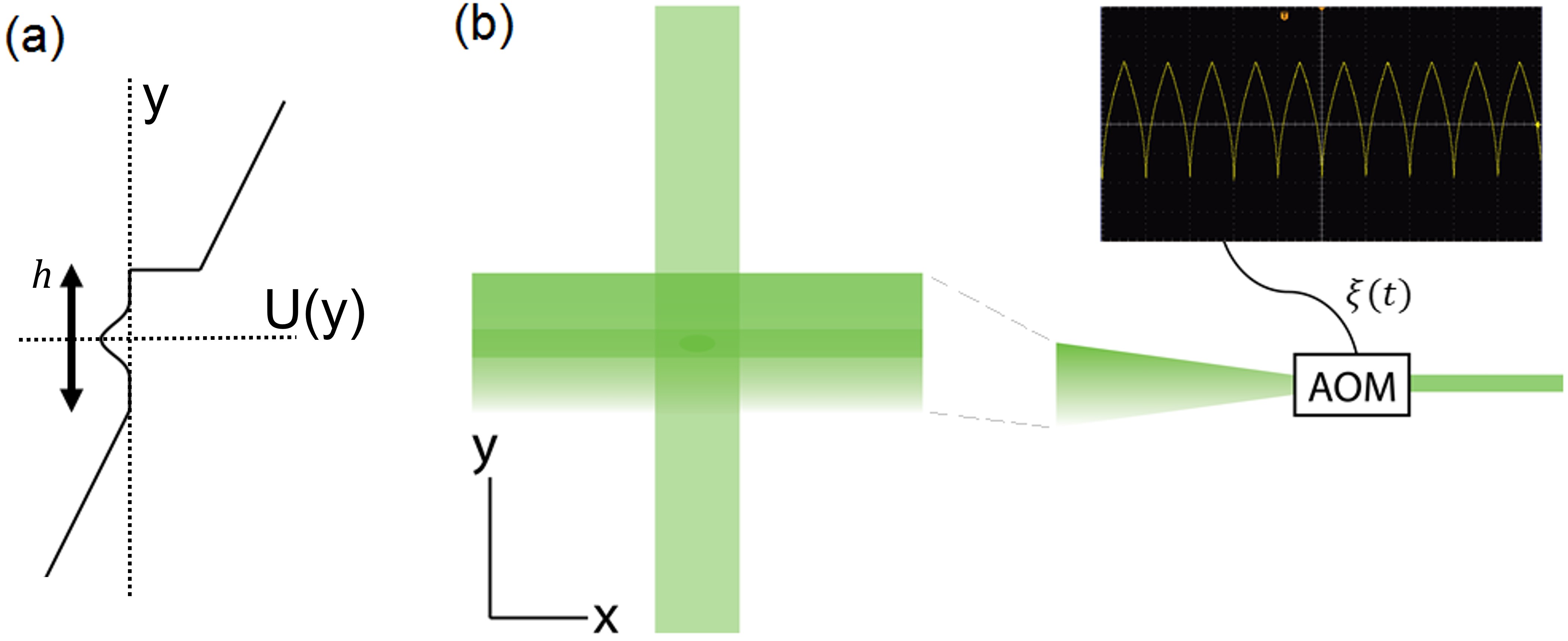}
	\caption{Implementation of a shaped optical beam for gravity compensation. (\textbf{a})~Model for total optical potential along the vertical axis, through the center of the optical trapping region. (\textbf{b}) Illustration of the ODT geometry along with the shaped optical beam. This beam is used to compensate the linear gravitational potential through use of an acousto-optic modulator supplied with the waveform, $\xi(t)$, as shown.}
	\label{fig:Figure2}
\end{figure}

The shaped beam used the same 532~nm light used for optical trapping. It was aligned to co-propagate with the horizontal ODT beam and---by means of an AOM---could be spatially modulated along the vertical axis (see Figure~\ref{fig:Figure2}). We designed the input waveform of the AOM such that the resultant time-averaged optical potential at the atoms would be linear over a region $h$, slightly larger than the trapping region. The function $\xi(t)=h\sqrt{\frac{\omega_p}{\pi}t}$ represents one half-period of this waveform and is therefore defined over $0\leq t<\pi$ with a full oscillation frequency, $\omega_p$. For this work $\omega_p=2\pi\times 4$~kHz, chosen to be much greater than the trap frequencies and less than the bandwidth of our electronics.

\begin{figure}[H]
	\centering
	\includegraphics[width=0.9\textwidth]{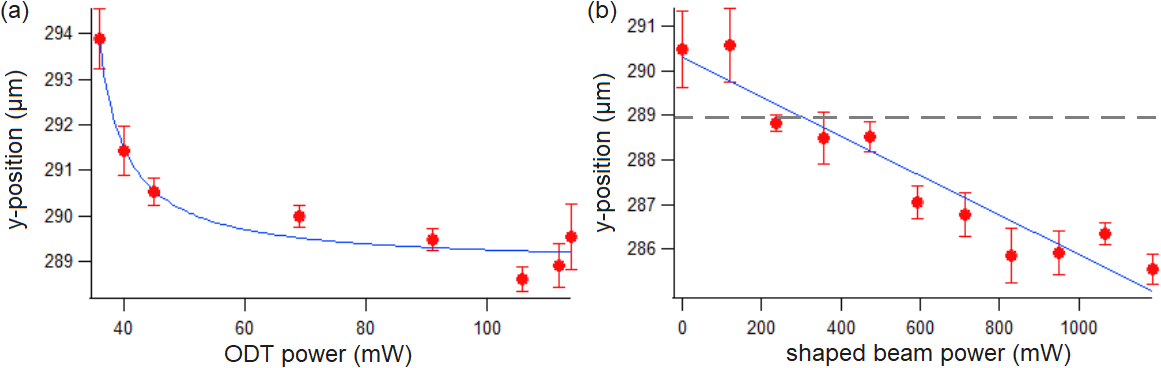}
	\caption{Optimization procedure for gravity compensation with a shaped optical beam. (\textbf{a})~Measured BEC position with absorption imaging, as the ODT depth is increased. The converging locations at higher depths indicate a reduction of the displacement due to the gravitational potential. The blue curve is a best-fit reciprocal function. (\textbf{b})~Introduction of the linear optical potential at low ODT depth. The BEC position has a nearly linear dependence on the shaped beam power as the net background potential gradient changes. The dashed line marks the asymptotic value of the reciprocal function shown in (a). The solid blue curve is a best-fit line to this data.}
	\label{fig:Figure3}
\end{figure}

The gradient of this optical potential was adjusted by changing the total power in the beam while maintaining a constant time-averaged beam intensity profile. This would alter the gradient of the net background potential in the trapping region (see Fig.~\ref{fig:Figure2}(a)), resulting in a displacement to the local minimum of the ODT. When this shift is zero, the linear optical potential is assumed to be properly compensating gravity. Thus, we first determined the ``zero point'' location by measuring the position of the atom cloud as a function of ODT depth in the absence of the new shaped optical beam. The best-fit curve for the data shown in Figure~\ref{fig:Figure3}(a) is a simple reciprocal function ($\propto$ 1/ODT power), derived from a harmonic approximation valid near the center of the Gaussian ODT intensity profile (see Eq.~\ref{eq:odtHO}). The convergent location observed at large ODT powers, corresponding to trap depths much higher than the gravitational potential variation across the trap, provided a benchmark value against which we could discern the effectiveness of the linear optical potential. The value is also marked in Fig.~\ref{fig:Figure3}(b) by the dashed line.

With the ODT returned to low depth (i.e. standard evaporation endpoint) we increased the total power in the shaped beam potential until the shift in the atoms' location was consistent with zero. This can be seen in Figure~\ref{fig:Figure3}(b) where the two lines cross. The best-fit curve for the data in this figure is a line, again as a result of approximating a harmonic trapping potential near the center of the ODT (see Eq.~\ref{eq:odtHO}). The optimal beam power from this procedure is then determined to be about 300~mW. Following this optimization, we used time-of-flight absorption imaging to measure the kinetic energy in the vertical direction to be $k_B \times (17\pm2)$~nK, a $42/17 \simeq 2.5$-fold improvement compared to without the gravity compensation beam. In terms of velocity spread, the reduction factor is $\sqrt{2.5} \simeq 1.6$.

%%%%%%%%%%%%%%%%%%%%%%%%%%%%%%%%%%%%%%%%%%
\section{Double Mach-Zehnder Interferometer}

We next report a demonstration of the first vertical Yb BEC interferometer. The geometry we use for this is a double Mach-Zehnder configuration, a design which is beneficial for precision measurement and sensing, since it suppresses vibration noise. In particular, the vertical double Mach-Zehnder interferometer is well-suited for gravity gradiometry applications \cite{mcgu02}.

Our implementation consists of four atom-optic elements: two splitting pulses, a mirror pulse, and a readout pulse. Each was implemented as a third-order Bragg pulse \cite{gupt01,goch19} with Gaussian $1/e$ full-width 54~$\mu$s in our vertical optical lattice with single-photon detuning, $\Delta=+3500\Gamma$. The typical peak lattice depth for a mirror pulse was $26\hbar\omega_{\rm rec}$ to satisfy the Bragg $\pi$--pulse condition, while the splitting and readout pulses had a peak depth of $14\hbar\omega_{\rm rec}$ and operated as $\pi/2$--pulses. Here $\omega_{\rm rec}=\hbar k_g^2/2m$ is the recoil frequency. The relative detuning of the lattice beams was chirped at the rate $\dot{\delta}=2gk_g$ for all pulses to account for the continuous Doppler shift of the falling atoms.

\begin{figure}[H]
	\centering
	\includegraphics[width=0.9\textwidth]{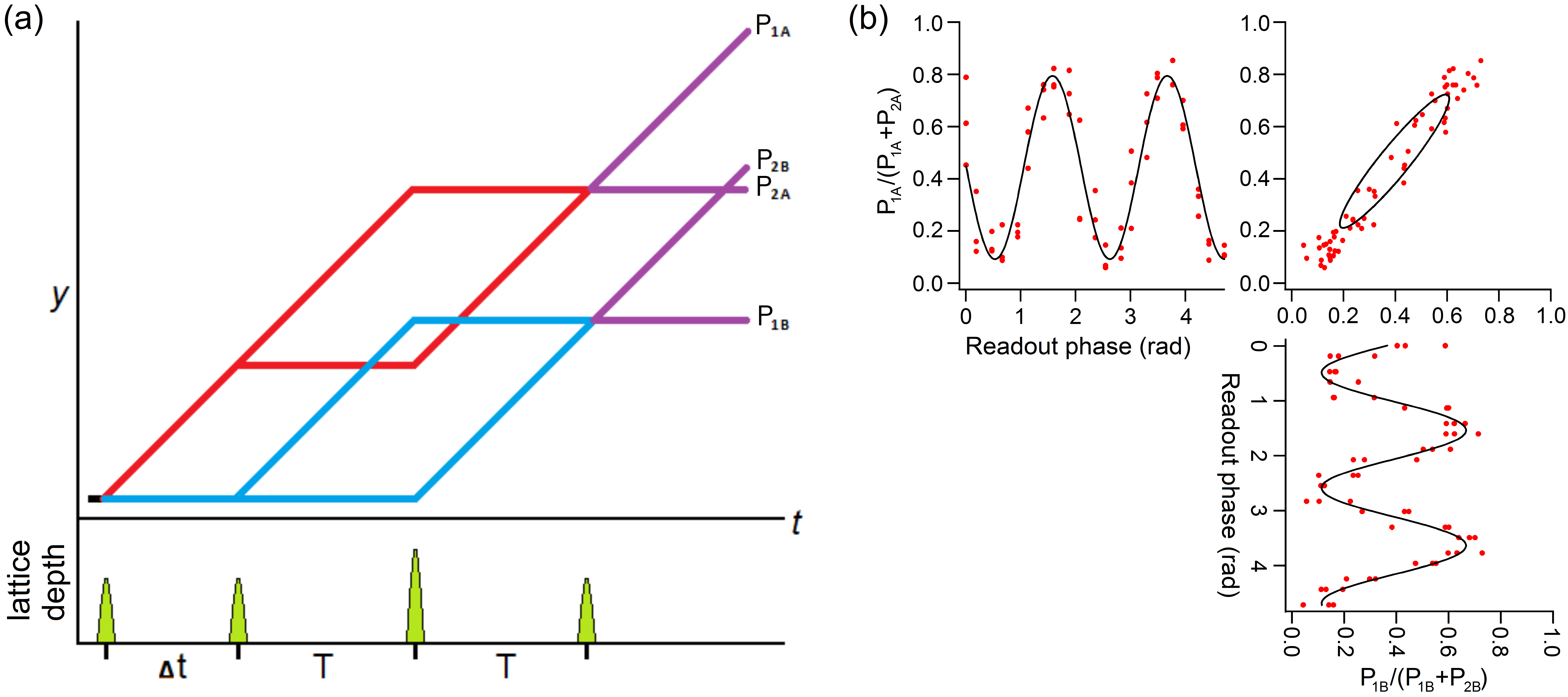}
	\caption{Double Mach-Zehnder interferometry. (\textbf{a})~Upper: Space-time diagram for a typical interferometer sequence shown in the accelerating frame of the free-falling BEC source such that the action of gravity is removed. The horizontal lines and the lines with finite slope correspond to interferometer paths separated in momentum by $6\hbar k_g$. Lower: Corresponding lattice pulse sequence. (\textbf{b})~Representative ellipse signal for a double Mach-Zehnder interferometer with $\Delta t=3$~ms and $T=0.25$~ms. The oscillating populations in the output ports are plotted versus the phase of the readout lattice for the two sub-interferometers, from which an ellipse can be observed in the correlation of the populations in the top right parametric plot. All black curves are best fit sinusoids or ellipses.}
	\label{fig:Figure4}
\end{figure}

The double Mach-Zehnder interferometer geometry is depicted in Figure~\ref{fig:Figure4}(a) in the accelerating frame of a falling atom cloud. To improve the efficiency of the momentum transfer within the interferometer, we apply an initial third-order Bragg $\pi$--pulse (not shown) to further narrow the width of the vertical velocity distribution. The two splitting pulses at the beginning of the interferometer are separated by a time $\Delta t$, chosen here to be 3~ms such that the interferometer paths are visually distinguishable in our absorption images. From these images we determine the relative populations in the output ports and observe interference fringes for each sub-interferometer, $A$ and $B$ (see Fig.~\ref{fig:Figure4}(b)). For long interferometer duration, however, physical vibrations effect an unknowable shift to the lattice phase, resulting in a reduction of fringe visibility for each sub-interferometer. Nonetheless, the differential interferometer phase for a double Mach-Zehnder, $\Delta\Phi=\Delta\phi_B-\Delta\phi_A$, is insensitive to vibration effects, which cancel out as a common mode phase shift. On the other hand, a finite differential phase can be generated from external forces such as a gravity gradient. This phase $\Delta\Phi$ can be observed by analyzing the correlation of sub-interferometer populations. As shown in Fig.~\ref{fig:Figure4}(b), the fractional population in an output port of one sub-interferometer when plotted against that in an output port of the other sub-interferometer traces out an ellipse whose eccentricity determines $\Delta \Phi$ \cite{fost02,stoc07,bert06,chio09}.

\begin{figure}[H]
	\centering
	\includegraphics[width=0.9\textwidth]{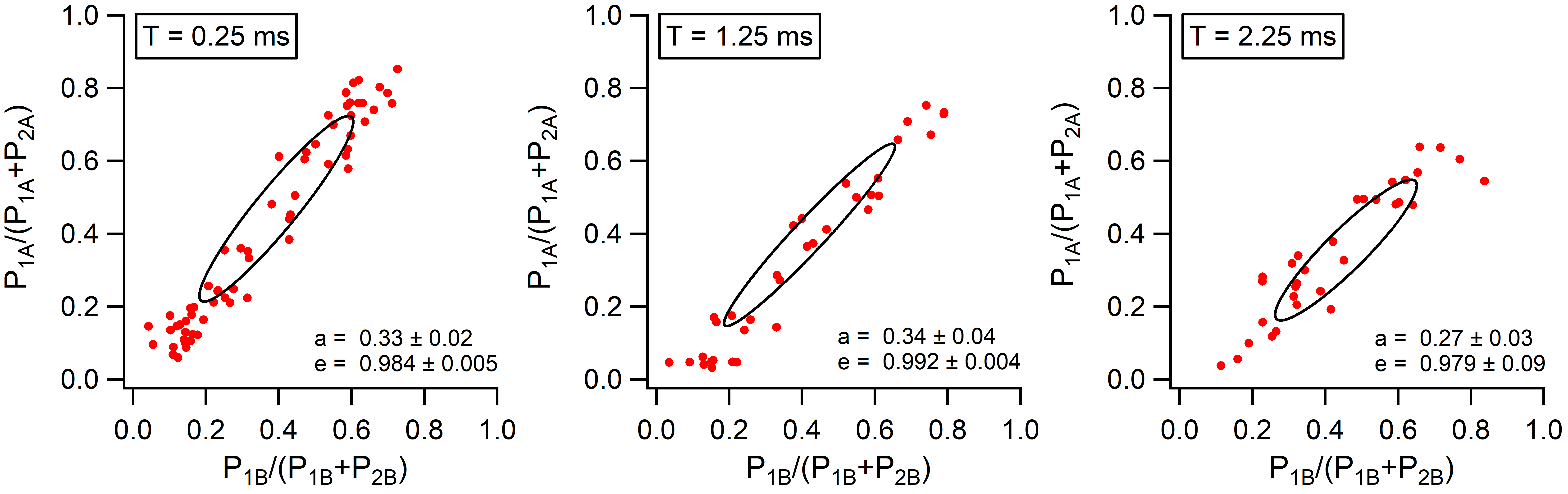}
	\caption{Ellipse signals at various values of $T$ for a double Mach-Zehnder interferometer. Black curves are best fit ellipses for each data set. These are characterized by $e$, the eccentricity of the ellipse and $a$, the length of the semi-major axis.}
	\label{fig:Figure5}
\end{figure}

To demonstrate robustness against vibrations, we extended the free evolution time within the interferometer, $T$, to values above the timescale of vibrations, which exist in our system at a bandwidth below 1~kHz. In this set of experiments, $T=0.25$~ms, 1.25~ms, and 2.25~ms, covering nearly one whole order of magnitude. The visibility at short times is as high as 80\% (see for $T=0.25\,$ms in Fig.~\ref{fig:Figure4}(b)), but drops with increasing $T$, and is consistent with zero by $T=2.25\,$ms. However, the ellipse traced out by the correlated populations is only marginally disturbed as shown in Fig.~\ref{fig:Figure5}. The fits (black curves in Fig.~\ref{fig:Figure5}) are obtained by first converting the data into polar coordinates, then performing a least-squares regression analysis using the function

\begin{equation}
    r=\frac{a(1-e^2)}{1+e\cos(\theta-\theta_0)}
\end{equation}

\noindent for an ellipse with one focus at the origin. In addition to the origin location, the given fit parameters describe the eccentricity of the ellipse, $e$, the length of the semi-major axis, $a$, and the rotation angle, $\theta_0$. The differential interferometer phase can be determined from the eccentricity by the relation \cite{fost02}

\begin{equation}
\Delta\Phi=\cos^{-1}\left(\frac{e^2}{2-e^2}\right)
\end{equation}

\noindent which is defined over one-quarter period.

The values of eccentricity returned by the fits are very close to 1, implying a differential phase close to zero. Indeed, the gravitational gradient for these parameters is expected to be negligibly small. The $\Delta \Phi$ corresponding to the measured $e \simeq 0.99$ is a few hundred mrad and may be due to atomic interactions. The calculated interaction energy for 3~ms expansion time and immediately before the first splitting $\pi/2$ pulse is 1.2~kHz. For a $10\%$ level difference in arm splitting this gives rise to few hundred mrad phase shift over the few millisecon timescale of the interferometer. This hypothesis is also consistent with the location of the center position of the ellipse deviating from the symmetric (0.5, 0.5) location at the 10\% level. This deviation is a manifestation of deviations at the same level from the $\pi/2$ condition for splitting pulses, and was independently verified by observing the correlation of the output port populations in the absence of the readout pulse. Further tuning of the $\pi/2$ pulses and longer expansion time to reduce the interaction strength, as is needed to scale-up the interferometer to larger times and enclosed areas, will make this differential phase contribution negligible.

%%%%%%%%%%%%%%%%%%%%%%%%%%%%%%%%%%%%%%%%%%
\section{Delta-kick Cooling of Yb for Vertical AI}

\subsection{Delta-kick Cooling}

We now report on another vertical cooling method for Yb---delta-kick cooling (DKC)---which shows even better performance than the gravity compensation beam technique described earlier. This cooling technique, also known as matter-wave lensing \cite{chu86,kova15,munt13}, has previously been demonstrated in alkali atoms, but not in spin-singlet atoms to the best of our knowledge. The process involves pulsing a parabolic attractive potential which may slow---or even halt---the expansion of the atom cloud in the corresponding dimension. In our system, this is applied as a pulsed optical potential at a time $t_o$ after the BEC had been released from the ODT. The potential is derived from our 556~nm laser with a detuning, $\Delta_{\rm DKC}=-4600\Gamma$, red-shifted from the ${}^1S_0\rightarrow{}^3P_1$ resonance. The DKC beam has a power of $\simeq 22\,$mW and is oriented horizontally, close to the horizontal ODT axis. It is focused with a waist size of $\simeq 150\,\mu$m at the location of the atoms. The DKC beam creates a transverse attractive potential for the atoms  proportional to the Gaussian shape of the beam, which is parabolic to lowest order near the beam center with effective harmonic angular frequency $\omega_{\rm DKC}$.

\begin{figure}[H]
	\centering
	\includegraphics[width=0.9\textwidth]{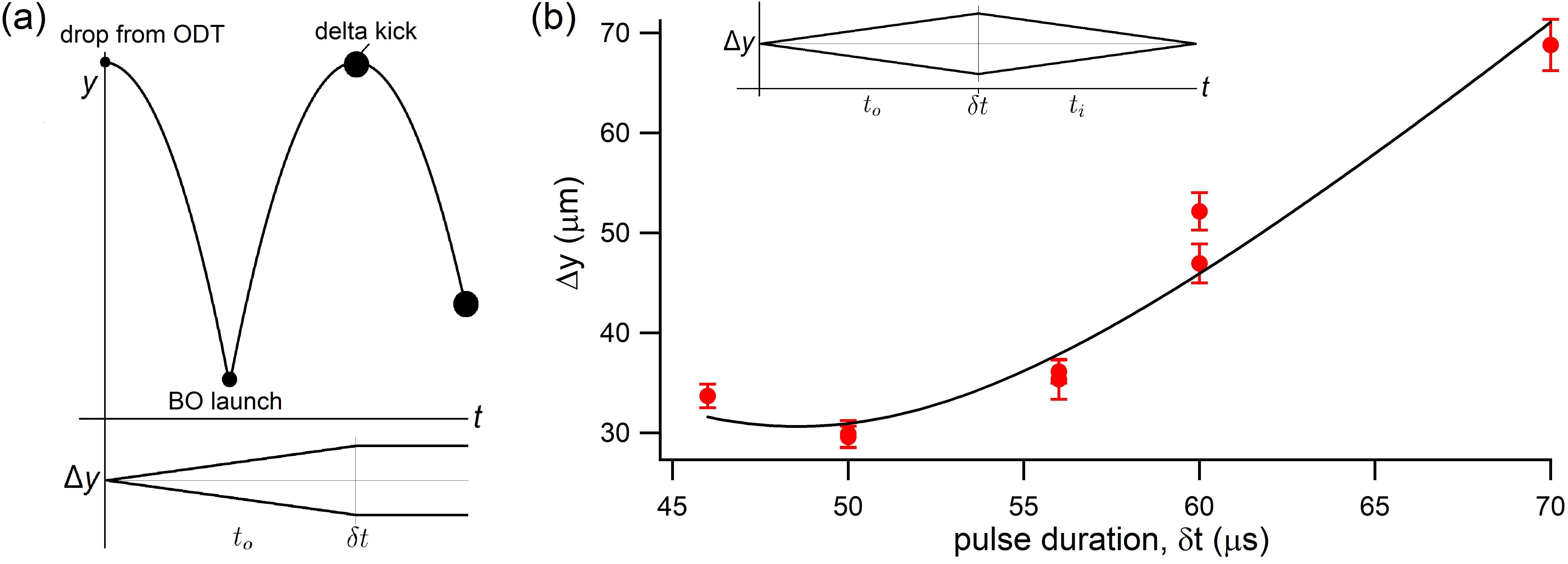}
	\caption{Delta-kick cooling and characterization sequences. (\textbf{a})~Space-time diagram depicting the trajectory of the expanding atom cloud. The DKC pulse of duration $\delta t$ is applied at time $t_o$ after release from the ODT to achieve collimation. (\textbf{b})~DKC characterization by measuring the size of a refocused atom cloud at time $t_i$ after the delta-kick. The black curve is a best-fit hyperbola, $(\Delta y)^2/(\Delta y_{\rm min})^2-(\delta t-\delta t_{\rm min})^2/C^2=1$, returning a minimum observable size $\Delta y_{\rm min}=30.6\pm2.0$~$\mu$m with a corresponding $\delta t_{\rm min}$ of $48.6\pm1.9$~$\mu$s. The DKC beam power here is approximately 22~mW.}
	\label{fig:Figure6}
\end{figure}

To minimize the variance of the BEC position over the duration of the DKC pulse, $\delta t$, the pulse was applied at the apex of the atoms' trajectory following a vertical fountain launch. To facilitate beam alignment, the launch was designed to place the apex at the location of the trapped BEC (see Fig.~\ref{fig:Figure6}(a)). The optical lattice had a depth of about $30\hbar\omega_{\rm rec}$ during the launch and a single-photon detuning $\Delta=-3500\Gamma$. For a typical experiment, the launch consisted of 30~Bloch oscillations effected by chirping the relative lattice detuning at a rate of 500~kHz/ms. With these launch parameters, the time for the atoms to reach the apex occurs $t_o=25.4$~ms after release from the ODT. For an optimal delta-kick we must simultaneously satisfy a thin lens criterion, $\delta t \ll t_o$, as well as the collimation condition, $\delta t \simeq 1/(\omega_{\rm DKC}^2t_o)$.   

We determined the capability of this cooling technique in our system from a characterization of our delta-kick lens and its effect on our atom source \cite{kova15}. For this experimental sequence, we applied the delta-kick at time $t_o$, tuned to refocus the atom cloud at time $t_i$ after the pulse, producing an image of the original object (BEC). By varying the pulse duration we ascertained a minimum observable spot size, $\Delta y_{\rm min} = 30.6\pm2.0$~$\mu$m, in the vertical dimension. Under collimation conditions, our system should have an upper bound on the minimum spread in the velocity distribution according to $\Delta v_{\rm bound}=\Delta y_{\rm min}/t_i$, where $\Delta v$ is the RMS velocity of the atoms after application of the lens. Through a simple relation based on the equipartition theorem, we can recast this as an upper bound on the temperature, $T_{\rm bound}=m(\Delta v_{\rm bound})^2/k_B$. From our measurement (see Fig.~\ref{fig:Figure6}(b)) and within the given experimental parameters, we determined an upper bound to the minimum attainable vertical temperature to be $T_{\rm bound}=4.0\pm0.5$~nK for our system.

\subsection{Ramsey Interferometry and Coherence Time Measurements}

\begin{figure}[H]
	\centering
	\includegraphics[width=1.0\textwidth]{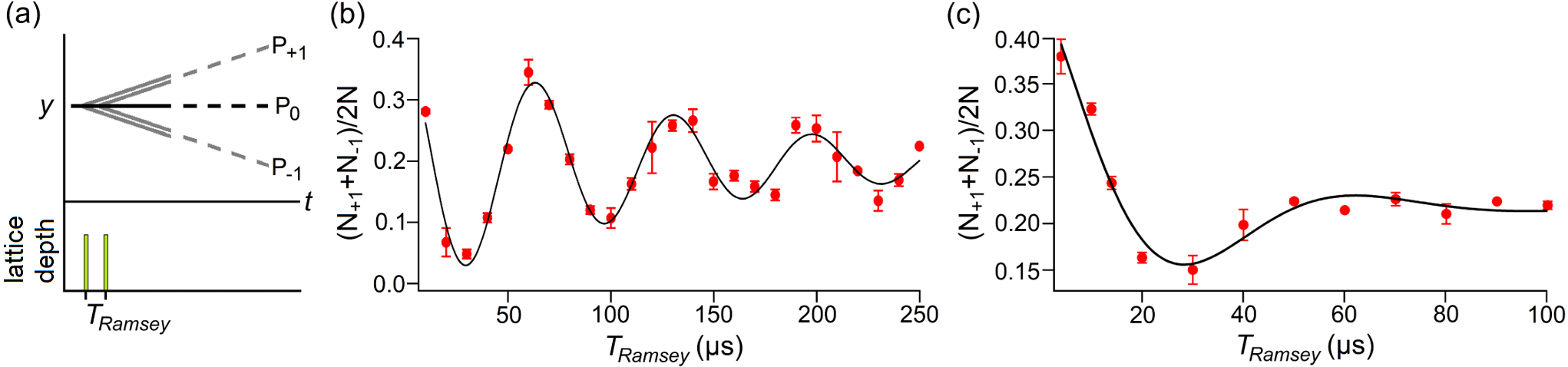}
	\caption{Characterization of DKC by coherence length measurements using Ramsey interferometry. (\textbf{a})~Space-time diagram of a momentum-space Ramsey interferometer composed of two Kapitza-Dirac pulses. Decaying oscillations of population amplitude, with (\textbf{b})~and without (\textbf{c})~delta-kick cooling. The fit function, $f(t)=Ae^{-t/\tau}\sin(4\omega_{\rm rec} t+\phi)+f_0$, returns Ramsey coherence times of $\tau=129\pm18$~$\mu$s and $23\pm3$~$\mu$s, respectively.}
	\label{fig:Figure7}
\end{figure}

As a secondary characterization of our delta-kick cooling lens, we analyzed the coherence length of the condensate before and after the DKC pulse. We incorporated a momentum-space Ramsey interferometer \cite{hagl99} into the sequence, which consisted of two low-amplitude 4~$\mu$s square Kapitza-Dirac pulses separated by a time, $T_{\rm Ramsey}$, as depicted in Figure~\ref{fig:Figure7}(a). The population of each momentum state was measured at long time of flight (i.e. enough time to spatially resolve the states) and oscillations were observed in the average fraction in the higher-momentum states, $(N_{+1}+N_{-1})/2N$. Here $N$ is the total atom number and $N_{\pm 1}$ is the atom number in the momentum state $\pm 2\hbar k$. Only at short times will there be observable fringes while the various wavefunction components retain sufficient spatial overlap. Thus, the envelope on these oscillations gives a Ramsey coherence time which is a measure of the coherence length of the atom source, with the two related through the velocity separation between the interfering states. Fitting the data to a sinusoid with the expected angular frequency $4\omega_{\rm rec}$ and an exponential envelope, we measure coherence times of $129\pm18$~$\mu$s and $23\pm3$~$\mu$s for sequences with and without delta-kick cooling, respectively. Since the coherence time scales with the inverse of the velocity spread, this indicates a reduction factor of $(129/23) \simeq 5.6$ in the vertical velocity spread. The reduction factor for kinetic energy in the vertical direction is then $(129/23)^2 \simeq 31$ below the previously mentioned $k_B \times 42\,$nK value, i.e., $k_B \times 1.3\,$nK. This is consistent with the upper bound on the kinetic energy $(1/2)k_B T_{\rm bound}$ of $k_B \times 2\,$nK, discussed in Section 5.1.

%%%%%%%%%%%%%%%%%%%%%%%%%%%%%%%%%%%%%%%%%%
\section{Conclusions}

In summary, we have investigated two methods to reduce the vertical velocity spread of a Yb BEC and applied them towards atom interferometry in vertical fountain geometries. We developed a gravity compensation optical potential which reduced the vertical velocity spread by a factor of 1.6, and employed this gain towards demonstrating a vertical double Mach-Zehnder interferometer. The implementation of delta-kick cooling reduced the vertical velocity spread by a factor of 5.6, which we measured using a Ramsey interferometer technique in an atomic fountain setup. These first demonstrations of vertical AI with Yb BECs reported here are performed with fountain times in the tens of milliseconds, and we expect that the reduction of vertical velocity spread achieved here should greatly benefit extensions to longer fountain times of hundreds of milliseconds.

Taken together, our results provide useful manipulation tools for ultracold Yb, an important atom for applications in fundamental physics and next generation time standards \cite{dzub18,hink13}, precision atom interferometry and sensing \cite{jami14,plot18,hart15,nied20,rour20}, quantum simulation \cite{taka03,seet21,paga14,scaz14,taka19} and quantum information science \cite{stoc08,sask19,cove19}. In particular, we expect the DKC methods to fruitfully impact future applications of Yb atom interferometers for precision sensing such as gravity gradiometry and tests of fundamental physics \cite{schl14,asen17,asen20,rour20}.

%%%%%%%%%%%%%%%%%%%%%%%%%%%%%%%%%%%%%%%%%%
\vspace{6pt} 

\funding{This work was supported by NSF Grant No. PHY-1707575.}

\reftitle{References}

% Please provide either the correct journal abbreviation (e.g. according to the “List of Title Word Abbreviations” http://www.issn.org/services/online-services/access-to-the-ltwa/) or the full name of the journal.
% Citations and References in Supplementary files are permitted provided that they also appear in the reference list here. 

%=====================================
% References, variant A: external bibliography
%=====================================
\externalbibliography{yes}
\bibliography{ifmrefs20}

\begin{thebibliography}{-------}
\providecommand{\natexlab}[1]{#1}

\bibitem[Bouchendira \em{et~al.}(2011)Bouchendira, Clade, Guellati-Khelifa,
  Nez, and Biraben]{bouc11}
Bouchendira, R.; Clade, P.; Guellati-Khelifa, S.; Nez, F.; Biraben, F.
\newblock New Determination of the Fine Structure Constant and Test of the
  Quantum Electrodynamics.
\newblock {\em Phys. Rev. Lett.} {\bf 2011}, {\em 106},~080801.

\bibitem[Mazzoni \em{et~al.}(2015)Mazzoni, Zhang, Del~Aguila, Salvi, Poli, and
  Tino]{mazz15}
Mazzoni, T.; Zhang, X.; Del~Aguila, R.; Salvi, L.; Poli, N.; Tino, G.
\newblock {Large-momentum-transfer Bragg interferometer with strontium atoms}.
\newblock {\em Phys. Rev. A} {\bf 2015}, {\em 92},~053619.

\bibitem[Hu \em{et~al.}(2017)Hu, Poli, Salvi, and Tino]{hupo17}
Hu, L.; Poli, N.; Salvi, L.; Tino, G.
\newblock Atom Interferometry with the Sr Optical Clock Transition.
\newblock {\em Phys. Rev. Lett.} {\bf 2017}, {\em 119},~263601.

\bibitem[Parker \em{et~al.}(2018)Parker, Yu, Estey, Zhong, and Muller]{park18}
Parker, R.H.; Yu, C.; Estey, B.; Zhong, W.; Muller, H.
\newblock {Measurement of the fine-structure constant as a test of the Standard
  Model}.
\newblock {\em Science} {\bf 2018}, {\em 360},~191.

\bibitem[Asenbaum \em{et~al.}(2017)Asenbaum, Overstreet, Kovachy, Brown, Hogan,
  and Kasevich]{asen17}
Asenbaum, P.; Overstreet, C.; Kovachy, T.; Brown, D.; Hogan, J.; Kasevich, M.
\newblock {Phase Shift in an Atom Interferometer due to Spacetime Curvature
  across its Wave Function}.
\newblock {\em Phys. Rev. Lett.} {\bf 2017}, {\em 118},~183602.

\bibitem[Jamison \em{et~al.}(2014)Jamison, Plotkin-Swing, and Gupta]{jami14}
Jamison, A.O.; Plotkin-Swing, B.; Gupta, S.
\newblock Advances in Precision Contrast Interferometry with Yb Bose-Einstein
  condensates.
\newblock {\em Phys. Rev. A} {\bf 2014}, {\em 90},~063606.

\bibitem[Rudolph \em{et~al.}(2020)Rudolph, Wilkason, Nantel, Swan, Holland,
  Jiang, Garber, Carman, and Hogan]{rudo20}
Rudolph, J.; Wilkason, T.; Nantel, M.; Swan, H.; Holland, C.; Jiang, Y.;
  Garber, B.; Carman, S.; Hogan, J.
\newblock {Large Momentum Transfer Clock Atom Interferometry on the 689 nm
  Intercombination Line of Strontium}.
\newblock {\em Phys. Rev. Lett.} {\bf 2020}, {\em 124},~083604.

\bibitem[Plotkin-Swing \em{et~al.}(2018)Plotkin-Swing, Gochnauer, McAlpine,
  Cooper, Jamison, and Gupta]{plot18}
Plotkin-Swing, B.; Gochnauer, D.; McAlpine, K.; Cooper, E.; Jamison, A.; Gupta,
  S.
\newblock {Three-Path Atom Interferometry with Large Momentum Separation}.
\newblock {\em Phys. Rev. Lett.} {\bf 2018}, {\em 121},~133201.

\bibitem[Gochnauer \em{et~al.}(2019)Gochnauer, McAlpine, Plotkin-Swing,
  Jamison, and Gupta]{goch19}
Gochnauer, D.; McAlpine, K.; Plotkin-Swing, B.; Jamison, A.; Gupta, S.
\newblock {Bloch-band picture for light-pulse atom diffraction and
  interferometry}.
\newblock {\em Phys. Rev. A.} {\bf 2019}, {\em 100},~043611.

\bibitem[McAlpine \em{et~al.}(2020)McAlpine, Gochnauer, and Gupta]{mcal20}
McAlpine, K.; Gochnauer, D.; Gupta, S.
\newblock {Excited-band Bloch oscillations for precision atom interferometry}.
\newblock {\em Phys. Rev. A.} {\bf 2020}, {\em 101},~023614.

\bibitem[Plotkin-Swing(2018)]{plot18thesis}
Plotkin-Swing, B.
\newblock Large Momentum Separation Matter Wave Interferometry.
\newblock {\em University of Washington PhD Thesis} {\bf 2018}.

\bibitem[Plotkin-Swing \em{et~al.}(2020)Plotkin-Swing, Wirth, Gochnauer,
  Rahman, McAlpine, and Gupta]{plot20}
Plotkin-Swing, B.; Wirth, A.; Gochnauer, D.; Rahman, T.; McAlpine, K.; Gupta,
  S.
\newblock {Crossed-beam slowing to enhance narrow-line ytterbium magneto-optic
  traps}.
\newblock {\em Rev. Sci. Instr.} {\bf 2020}, {\em 91},~093201.

\bibitem[Castin and Dum(1996)]{cast96}
Castin, Y.; Dum, R.
\newblock {Bose-Einstein Condensates in Time Dependent Traps}.
\newblock {\em Phys. Rev. Lett.} {\bf 1996}, {\em 77},~5315.

\bibitem[Jamison \em{et~al.}(2011)Jamison, Kutz, and Gupta]{jami11}
Jamison, A.O.; Kutz, J.N.; Gupta, S.
\newblock Atomic Interactions in Precision Interferometry Using Bose-Einstein
  Condensates.
\newblock {\em Phys. Rev. A.} {\bf 2011}, {\em 84},~043643.

\bibitem[Shibata \em{et~al.}(2020)Shibata, Ikeda, Suzuki, and Hirano]{shib20}
Shibata, K.; Ikeda, H.; Suzuki, R.; Hirano, T.
\newblock {Compensation of gravity on cold atoms by a linear optical
  potential}.
\newblock {\em Phys. Rev. Research} {\bf 2020}, {\em 2},~013068.

\bibitem[McGuirk \em{et~al.}(2002)McGuirk, Foster, Fixler, Snadden, and
  Kasevich]{mcgu02}
McGuirk, J.M.; Foster, G.T.; Fixler, J.B.; Snadden, M.J.; Kasevich, M.A.
\newblock {Sensitive absolute-gravity gradiometry using atom interferometry}.
\newblock {\em Phys. Rev. A} {\bf 2002}, {\em 65},~033608.

\bibitem[Gupta \em{et~al.}(2001)Gupta, Leanhardt, Cronin, and
  Pritchard]{gupt01}
Gupta, S.; Leanhardt, A.E.; Cronin, A.D.; Pritchard, D.E.
\newblock {Coherent Manipulation of Atoms with Standing Light Waves}.
\newblock {\em Cr. Acad. Sci. IV-Phys} {\bf 2001}, {\em 2},~479.

\bibitem[Foster \em{et~al.}(2002)Foster, Fixler, McGuirk, and Kasevich]{fost02}
Foster, G.T.; Fixler, J.B.; McGuirk, J.M.; Kasevich, M.A.
\newblock {Method of phase extraction between coupled atom interferometers
  using ellipse-specific fitting}.
\newblock {\em Optics Letters} {\bf 2002}, {\em 27},~951.

\bibitem[Stockton \em{et~al.}(2007)Stockton, Wu, and Kasevich]{stoc07}
Stockton, J.K.; Wu, X.; Kasevich, M.A.
\newblock Bayesian estimation of differential interferometer phase.
\newblock {\em Phys. Rev. A} {\bf 2007}, {\em 76},~033613.

\bibitem[Bertoldi \em{et~al.}(2006)Bertoldi, Lamporesi, Cacciapuoti,
  de~Angelis, Fattori, Petelski, Peters, Prevedelli, Stuhler, and Tino]{bert06}
Bertoldi, A.; Lamporesi, G.; Cacciapuoti, L.; de~Angelis, M.; Fattori, M.;
  Petelski, T.; Peters, A.; Prevedelli, M.; Stuhler, J.; Tino, G.
\newblock Atom interferometry gravity-gradiometer for the determination of the
  Newtonian gravitational constant G.
\newblock {\em Eur. Phys. J. D} {\bf 2006}, {\em 40},~271--279.

\bibitem[Chiow \em{et~al.}(2009)Chiow, Herrmann, Chu, and Muller]{chio09}
Chiow, S.; Herrmann, S.; Chu, S.; Muller, H.
\newblock Noise-Immune Conjugate Large-Area Atom Interferometers.
\newblock {\em Phys. Rev. Lett.} {\bf 2009}, {\em 103},~050402.

\bibitem[Chu \em{et~al.}(1986)Chu, Bjorkholm, Ashkin, Gordon, and
  Hollberg]{chu86}
Chu, S.; Bjorkholm, J.; Ashkin, A.; Gordon, J.; Hollberg, L.
\newblock Proposal for optically cooling atoms to temperatures of the order of
  $10^{-6}$~K.
\newblock {\em Optics Letters} {\bf 1986}, {\em 11},~73.

\bibitem[Kovachy \em{et~al.}(2015)Kovachy, Hogan, Sugarbaker, Dickerson,
  Donnelly, Overstreet, and Kasevich]{kova15}
Kovachy, T.; Hogan, J.; Sugarbaker, A.; Dickerson, S.; Donnelly, C.;
  Overstreet, C.; Kasevich, M.
\newblock Matter Wave Lensing to Picokelvin Temperatures.
\newblock {\em Phys. Rev. Lett.} {\bf 2015}, {\em 114},~143004.

\bibitem[Muntinga \em{et~al.}(2013)Muntinga, Ahlers, Krutzik, Wenzlawski,
  Arnold, Becker, Bongs, Dittus, Duncker, Gaaloul, Gherasim, Giese, Grzeschik,
  Hansch, Hellmig, Herr, Herrmann, Kajari, Kleinert, Lammerzahl,
  Lewoczko-Adamczyk, Malcolm, Meyer, Nolte, Peters, Popp, Reichel, Roura,
  Rudolph, Schiemangk, Schneider, Seidel, Sengstock, Tamma, Valenzuela, Vogel,
  Walser, Wendrich, Windpassinger, Zeller, van Zoest, Ertmer, Schleich, and
  Rasel]{munt13}
Muntinga, H.; Ahlers, H.; Krutzik, M.; Wenzlawski, A.; Arnold, S.; Becker, D.;
  Bongs, K.; Dittus, H.; Duncker, H.; Gaaloul, N.; Gherasim, C.; Giese, E.;
  Grzeschik, C.; Hansch, T.; Hellmig, O.; Herr, W.; Herrmann, S.; Kajari, E.;
  Kleinert, S.; Lammerzahl, C.; Lewoczko-Adamczyk, W.; Malcolm, J.; Meyer, N.;
  Nolte, R.; Peters, A.; Popp, M.; Reichel, J.; Roura, A.; Rudolph, J.;
  Schiemangk, M.; Schneider, M.; Seidel, S.; Sengstock, J.; Tamma, V.;
  Valenzuela, T.; Vogel, A.; Walser, R.; Wendrich, T.; Windpassinger, P.;
  Zeller, W.; van Zoest, T.; Ertmer, W.; Schleich, W.; Rasel, E.
\newblock {Interferometry with Bose-Einstein Condensates in Microgravity}.
\newblock {\em Phys. Rev. Lett.} {\bf 2013}, {\em 110},~093602.

\bibitem[Hagley \em{et~al.}(1999)Hagley, Deng, Kozuma, Trippenbach, Band,
  Edwards, Doery, Julienne, Helmerson, Rolston, and Phillips]{hagl99}
Hagley, E.W.; Deng, L.; Kozuma, M.; Trippenbach, M.; Band, Y.B.; Edwards, M.;
  Doery, M.; Julienne, P.S.; Helmerson, K.; Rolston, S.L.; Phillips, W.D.
\newblock Measurement of the Coherence of a Bose-Einstein Condensate.
\newblock {\em Phys. Rev. Lett.} {\bf 1999}, {\em 83},~3112.

\bibitem[Dzuba \em{et~al.}(2018)Dzuba, Flambaum, and Schiller]{dzub18}
Dzuba, V.; Flambaum, V.; Schiller, S.
\newblock {Testing physics beyond the standard model through additional clock
  transitions in neutral ytterbium}.
\newblock {\em Phys. Rev. A} {\bf 2018}, {\em 98},~022501.

\bibitem[Hinkley \em{et~al.}(2013)Hinkley, Sherman, Phillips, Schioppo, Lemke,
  Beloy, Pizzocaro, Oates, and Ludlow]{hink13}
Hinkley, N.; Sherman, J.; Phillips, N.; Schioppo, M.; Lemke, N.; Beloy, K.;
  Pizzocaro, M.; Oates, C.; Ludlow, A.
\newblock An Atomic Clock With $10^{-18}$ Instability.
\newblock {\em Science} {\bf 2013}, {\em 341},~1215--1218.

\bibitem[Hartwig \em{et~al.}(2015)Hartwig, Abend, Schubert, Schlippert, Ahlers,
  Posso-Trujillo, Gaaloul, Ertmer, and Rasel]{hart15}
Hartwig, J.; Abend, S.; Schubert, S.; Schlippert, D.; Ahlers, H.;
  Posso-Trujillo, K.; Gaaloul, N.; Ertmer, W.; Rasel, E.
\newblock {Testing the universality of free fall with rubidium and ytterbium in
  a very large baseline atom interferometer}.
\newblock {\em New J. Phys.} {\bf 2015}, {\em 17},~035011.

\bibitem[Niederriter \em{et~al.}(2020)Niederriter, Schlupf, and
  Hamilton]{nied20}
Niederriter, R.; Schlupf, C.; Hamilton, P.
\newblock {Cavity probe for real-time detection of atom dynamics in an optical
  lattice}.
\newblock {\em Phys. Rev. A} {\bf 2020}, {\em 102},~051301(R).

\bibitem[Roura \em{et~al.}(2020)Roura, Schubert, Schlippert, and Rasel]{rour20}
Roura, A.; Schubert, C.; Schlippert, D.; Rasel, E.
\newblock Measuring gravitational time dilation with delocalized quantum
  superpositions.
\newblock {\em Arxiv} {\bf 2020}, p. 2010.11156.

\bibitem[Takasu \em{et~al.}(2002)Takasu, Maki, Komori, Takano, Honda, Kumakura,
  Yabuzaki, and Takahashi]{taka03}
Takasu, Y.; Maki, K.; Komori, K.; Takano, T.; Honda, K.; Kumakura, M.;
  Yabuzaki, T.; Takahashi, Y.
\newblock {Spin-singlet Bose-Einstein Condensation of Two-Electron Atoms}.
\newblock {\em Phys. Rev. Lett.} {\bf 2002}, {\em 91},~040404.

\bibitem[See~Toh \em{et~al.}(2021)See~Toh, McCormick, Tang, Su, Luo, Zhang, and
  Gupta]{seet21}
See~Toh, J.; McCormick, K.; Tang, X.; Su, Y.; Luo, X.; Zhang, C.; Gupta, S.
\newblock {Observation of Many-body Dynamical Delocalization in a Kicked
  Ultracold Gas}.
\newblock {\em Arxiv} {\bf 2021}, p. 2106.13773.

\bibitem[Pagano \em{et~al.}(2014)Pagano, Mancini, Cappellini, Lombardi,
  Schäfer, Hu, Liu, Catani, Sias, Inguscio, and Fallani]{paga14}
Pagano, G.; Mancini, M.; Cappellini, G.; Lombardi, P.; Schäfer, F.; Hu, H.;
  Liu, X.; Catani, J.; Sias, C.; Inguscio, M.; Fallani, L.
\newblock A one-dimensional liquid of fermions with tunable spin.
\newblock {\em Nat. Phys.} {\bf 2014}, {\em 10},~198--201.

\bibitem[Scazza \em{et~al.}(2014)Scazza, Hofrichter, Höfer, De~Groot, Bloch,
  and Fölling]{scaz14}
Scazza, F.; Hofrichter, C.; Höfer, M.; De~Groot, P.; Bloch, I.; Fölling, S.
\newblock Observation of two-orbital spin-exchange interactions with ultracold
  SU(N)-symmetric fermions.
\newblock {\em Nat. Phys.} {\bf 2014}, {\em 10},~779--784.

\bibitem[Ono \em{et~al.}(2019)Ono, Kobayashi, Amano, Sato, and
  Takahashi]{taka19}
Ono, K.; Kobayashi, J.; Amano, Y.; Sato, K.; Takahashi, Y.
\newblock Antiferromagnetic interorbital spin-exchange interaction of
  $^{171}$Yb.
\newblock {\em Phys. Rev. A} {\bf 2019}, {\em 99},~032707.

\bibitem[Stock \em{et~al.}(2008)Stock, Babcock, Raizen, and Sanders]{stoc08}
Stock, R.; Babcock, N.; Raizen, M.; Sanders, B.
\newblock Entanglement of group-II-like atoms with fast measurement for quantum
  information processing.
\newblock {\em Phys. Rev. A} {\bf 2008}, {\em 78},~022301.

\bibitem[Saskin \em{et~al.}(2019)Saskin, Wilson, Grinkemeyer, and
  Thompson]{sask19}
Saskin, S.; Wilson, J.; Grinkemeyer, B.; Thompson, J.
\newblock Narrow-Line Cooling and Imaging of Ytterbium Atoms in an Optical
  Tweezer Array.
\newblock {\em Phys. Rev. Lett.} {\bf 2019}, {\em 122},~143002.

\bibitem[Covey \em{et~al.}(2019)Covey, Sipahigil, Szoke, Sinclair, Endres, and
  Painter]{cove19}
Covey, J.; Sipahigil, A.; Szoke, S.; Sinclair, N.; Endres, M.; Painter, O.
\newblock Telecom-Band Quantum Optics with Ytterbium Atoms and Silicon
  Nanophotonics.
\newblock {\em Phys. Rev. App.} {\bf 2019}, {\em 11},~034044.

\bibitem[Schlippert \em{et~al.}(2014)Schlippert, Hartwig, Albers, Richardson,
  Schubert, Roura, Schleich, Ertmer, and Rasel]{schl14}
Schlippert, D.; Hartwig, J.; Albers, H.; Richardson, L.; Schubert, C.; Roura,
  A.; Schleich, W.; Ertmer, W.; Rasel, E.
\newblock {Quantum Test of the Universality of Free Fall}.
\newblock {\em Phys. Rev. Lett.} {\bf 2014}, {\em 112},~203002.

\bibitem[Asenbaum \em{et~al.}(2020)Asenbaum, Overstreet, Kim, Curti, and
  Kasevich]{asen20}
Asenbaum, P.; Overstreet, C.; Kim, M.; Curti, J.; Kasevich, M.
\newblock {Atom-Interferometric Test of the Equivalence Principle at the
  $10^{-12}$ Level}.
\newblock {\em Phys. Rev. Lett.} {\bf 2020}, {\em 125},~191101.

\end{thebibliography}

%=====================================
% References, variant B: internal bibliography
%=====================================
%\begin{thebibliography}{999}
% Reference 1
%\bibitem[Author1(year)]{ref-journal}
%Author1, T. The title of the cited article. {\em Journal Abbreviation} {\bf 2008}, {\em 10}, 142--149.
% Reference 2
%\bibitem[Author2(year)]{ref-book}
%Author2, L. The title of the cited contribution. In {\em The Book Title}; Editor1, F., Editor2, A., Eds.; Publishing House: City, Country, 2007; pp. 32--58.
%\end{thebibliography}

% The following MDPI journals use author-date citation: Arts, Econometrics, Economies, Genealogy, Humanities, IJFS, JRFM, Laws, Religions, Risks, Social Sciences. For those journals, please follow the formatting guidelines on http://www.mdpi.com/authors/references
% To cite two works by the same author: \citeauthor{ref-journal-1a} (\citeyear{ref-journal-1a}, \citeyear{ref-journal-1b}). This produces: Whittaker (1967, 1975)
% To cite two works by the same author with specific pages: \citeauthor{ref-journal-3a} (\citeyear{ref-journal-3a}, p. 328; \citeyear{ref-journal-3b}, p.475). This produces: Wong (1999, p. 328; 2000, p. 475)

%%%%%%%%%%%%%%%%%%%%%%%%%%%%%%%%%%%%%%%%%%
%% optional
%\sampleavailability{Samples of the compounds ...... are available from the authors.}

%% for journal Sci
%\reviewreports{\\
%Reviewer 1 comments and authors’ response\\
%Reviewer 2 comments and authors’ response\\
%Reviewer 3 comments and authors’ response
%}

%%%%%%%%%%%%%%%%%%%%%%%%%%%%%%%%%%%%%%%%%%
\end{document}